\documentclass[a4paper,12pt]{article}
\usepackage[utf8]{inputenc}
\usepackage{physics}

\title{String Theory, Loop Quantum Gravity \\ and Eternalism}

\usepackage{url}
\usepackage{natbib}
\usepackage{amsmath}
\usepackage{amsfonts}
\usepackage[font={small}]{caption}
\usepackage[margin=0.8in]{geometry}
\usepackage{float}
\usepackage{url}
\usepackage{braket}
\usepackage{marginnote}

\author{Baptiste Le Bihan}
\date{\textit{forthcoming in \\ the European Journal for Philosophy of Science}}
\begin{document}


\maketitle

\abstract{Eternalism, the view that what we regard locally as being located in the past, the present and the future equally exists, is the best ontological account of temporal existence in line with special and general relativity. However, special and general relativity are not fundamental theories and several research programs aim at finding a more fundamental theory of quantum gravity weaving together all we know from relativistic physics and quantum physics. Interestingly, some of these approaches assert that time is not fundamental. If time is not fundamental, what does it entail for eternalism and the standard debate over existence in time? First, I will argue that the non-fundamentality of time to be found in string theory entails standard eternalism. Second, I will argue that the non-fundamentality of time to be found in loop quantum gravity entails atemporal eternalism, namely a novel position in the spirit of standard eternalism.}


\tableofcontents
\section{Introduction}

Some approaches to quantum gravity suggest that \emph{spacetime} (or at least some of its substantial features) is not fundamental.\footnote{For references in the philosophical literature, see e.g. \cite{HuggettWuthrich2013}, \cite{Matsubara2017}, \cite{BLBNL}, \cite{LW}, \citeauthor{baron} (forthcoming) and \citeauthor{HUWU} (forthcoming).} Quantum gravity is the name of a collection of research programs which aim at reconciling general relativity with quantum physics, our two most fundamental, and empirically confirmed, physical theories at the moment. On the one hand, general relativity (hereafter: ``GR'') is an excellent theory to describe gravitation---in particular at large scales. According to the standard reading of GR, gravitation is not a force, but the manifestation of the geometry of spacetime. However, GR does not take into account quantum effects and cannot be fully combined with quantum field theory. On the other hand, quantum field theory does an excellent work at describing quantum effects and fundamental forces---in particular at very small scales. However, it does not include a description of gravitation. As a result, we lack a theory giving us resources to describe \textit{phenomena} where both gravitational and quantum effects enter the picture: \emph{black holes}, the \emph{earliest times} after the big bang, and the \emph{structure of spacetime} itself.\footnote{This theory of quantum gravity might turn out to be a final theory of everything but it certainly does not have to be so. Certain research programs in quantum gravity are very explicit they are not looking for a final theory but only, more modestly, for a unified account of quantum gravity.}



Although these programs are highly speculative in nature, they are rich of potential repercussions for classical disputes in metaphysics. This article examines the possible consequences of the non-fundamentality of spacetime, and in particular of time, to be found in some of these approaches on the debate over \emph{existence in time}. \emph{Eternalism} is the view that past, present and future entities exist \textit{simpliciter},\footnote{This is a rough characterisation. Taking the general theory of relativity seriously implies defining eternalism without using the notions of past, present and future in the following way: all natural things exist independently of their location in the network of spatio-temporal or causal relations.} \emph{presentism} the view that only present entities exist \textit{simpliciter}, and \emph{no-futurism}, the view that past and present entities exist \textit{simpliciter} contrary to future ones. I will argue that the non-fundamentality of time, if borne out, will entail the truth of standard eternalism or a novel position: \emph{atemporal eternalism}, namely the view that any proper part of the natural world exists \textit{simpliciter} and that the material content of the natural world does not depend on any particular location in it.\footnote{The expression ``atemporal eternalism'' is already used in the philosophy of religion and refers to the view that God exists beyond time (see e.g. \citealt{Morris1991-MOROIO}). I will still use this expression as it describes well the content of the position; arguably, there is no risk to conflate the two views.}

Claims about the non-fundamentality of spacetime may strike the reader as rather strange since the view that things exist in space and time---or spacetime---is deeply rooted in our conceptual apparatus. This strangeness, combined with the lack of empirical support of those research programs and the fact that they still lack a comprehensive formulation, explains why the metaphysician can be (and has been) inclined to rather seek philosophical insight from contemporary physics by looking at our most fundamental---and empirically confirmed---theories. 

However, as \citeauthor{McKenzie} (forthcoming) recently argued, a naturalistic metaphysics based on this physics is at risk of being completely wrong since GR and quantum field theory are at best approximately true.\footnote{The interest of quantum gravity for metaphysics has been noted on several occasions; see e.g. \cite{HuggettWuthrich2013}, \cite{Muntean2015} and \cite{norton2017incubating}.}  In response to McKenzie's challenge, I propose to shift the focus from general relativity and quantum field theory to quantum gravity. Indeed, I take McKenzie's argument not to doom naturalized metaphysics, but rather to establish that naturalized metaphysics may not rely on empirically confirmed physics only since this physics is made of mutually exclusive, non-fundamental, parts. Metaphysics cannot rely on philosophical constraints inherited from two mutually exclusive physical theories and ignore the speculative models or fragments of theories which aim at combining the two views. But is there any reliable epistemological road from quantum gravity to metaphysics? After all, the very plurality of available approaches and their lack of a complete formulation seem to bar the road to any straightforward answer to metaphysical debates such as the one for existence in time.

In response, I believe there are two promising strategies to extract metaphysical know-ledge from the constellation of research programs in quantum gravity. The first strategy, that I will refer to as the \emph{ambitious epistemological path}, starts with the realization that some features appear in all standard approaches to quantum gravity. This general presence gives a high justification to some claims about the nature of the world. For instance, as far as non-locality\footnote{Non-locality refers here to the property ascribed to physical systems in response to the violation of Bell inequalities.} appears in all theories of quantum mechanics and in quantum field theories, and as far as quantum gravity does not make non-locality go away, this feature inherits some substantial justification from contemporary physics as a whole.\footnote{This of course does not tell us yet what is the right \emph{metaphysical interpretation} of non-locality. One might argue for instance that it entails \textit{priority monism} (see \citealp{ismael2016quantum}) or, alternatively, a form of \emph{coherentism} as defended by \citeauthor{Calosi} (forthcoming).} Or, to take another example, the fact that a local split between two structures corresponding more or less to space and time is to be found in all empirically confirmed physical theories and in all approaches to quantum gravity establishes that our world can be locally split in a spatial or quasi-spatial structure on the one hand, and a temporal or quasi-temporal structure on the other hand (cf. \citealt{BLBNL}). One might generalize the situation by noting as Rovelli and Vidotto do:

\begin{quote}
As far as we know today, all that exist in nature are generally covariant
quantum fields. \citep[19]{RovelliVidotto}
\end{quote}

Understanding exactly the ontology behind a mathematical description in terms of generally covariant quantum fields is a tremendously difficult task. My point is not to defend a particular ontological interpretation here, but merely to mention that looking at research programs in quantum gravity allows us to take a step back in order to evaluate which part of the quantum and relativistic revolutions are---beyond reasonable doubt---to be preserved in the future.
But this first road can hardly be taken as of today when it comes to the non-fundamentality of time. As Callender puts it:

\begin{quote}
Strategies for pursuing quantum gravity split in two, as do corresponding attitudes about time. If one feels that the quantum revolution provides the firmer or more promising foundation, then—as in superstring theory—one begins there and hence starts with a full-blooded time. If by contrast one believes that relativity provides the better starting point, then—as in loop quantum gravity—one begins with a theory wherein time is already demoted in some sense. In both cases one knows that the final product will be significantly different from relativity or quantum mechanics, and one needn’t end where one begins. Nonetheless, these initial decisions about time tend to have serious ramifications later. \emph{Because the field is so split, one can find confirmation of almost anything one thinks about time}. \cite[99, my emphasis]{callender2017makes}
\end{quote}

As we shall see, and although string theory \emph{starts} with a full-blooded time, it is not clear whether string theory will posit the existence of spacetime or time at the end of its journey: different opinions prevail within the string community. But luckily there is another, less ambitious, road one might examine when one aims at getting ontological insight from contemporary physics. Instead of looking for assumptions common to all approaches to quantum gravity, one might look at the \emph{specific ontological commitments} of one or several research programs, and think about the consequences for metaphysics if this, or one of these, approach(es) turns out to be right. For instance, and simplifying, one might take the disappearance of spacetime in \textit{string theory} to amount to the disappearance of at least space, because of the phenomenon of duality\footnote{See e.g. \citet{HuggettDualities}, \citet[3-4]{Matsubara2017}, \cite{BLBJR} and \citeauthor{ButterfieldPBS} (forthcoming).} and, perhaps as we shall see below, of time. In other approaches, on the contrary, the ontology we have seems closer to the denial of the existence of time, but preserves space---as in \textit{Barbour's ontology of time capsules}.\footnote{See \cite{barbour2001end} and \cite{Butterfield2002-BUTTEO}, \cite{healey2002can}, \cite*{baron2010timeless}, \cite{BLB2015unrealities} for philosophical discussion.} Or it could be that both space and time are evacuated from the picture as in \textit{loop quantum gravity}, as we shall see later on. 

This second strategy, that I will call the \emph{modest epistemological path}, delivers only \emph{conditional knowledge}: \emph{if} loop quantum gravity, or another approach denying reality to time were to be borne out, \emph{then} we would have good evidence for a particular metaphysical claim. Although this approach might seem a bit disappointing, conditional knowledge remains knowledge and this project merits to be pursued. Furthermore, the building of ontological frameworks consistent with these approaches might even help clarifying their development---indeed, determining which ontologies are consistent or even suggested by quantum gravity remains the object of ongoing research.\footnote{The question of how quantum gravity relates to existence in time is not new. \cite{monton2006presentism} acknowledging that presentism is inconsistent with special relativity argued that presentism could in principle be recovered from a theory of quantum gravity; a claim that was criticized by \cite{wuthrich2010no}. The evidence that Wüthrich is right against Monton is compelling. My goal in this paper is, starting from there, to discuss eternalism and how it fits with the non-fundamentality of time.} 







In section 2, I introduce the debate about existence in time. In sections 3 and 4, I describe the non-fundamentality of time in two approaches to quantum gravity, namely the \textit{two-dimensional field theory}---which is an approach to \textit{string theory}---and \textit{loop quantum gravity}. In section 5, I argue that this non-fundamentality entails eternalism. In section 6, I present several metaphysical interpretations of the claim that time disappears at the fundamental level, some realist interpretations offering the possibility to escape eternalism. In section 7, however, I examine the fallout of those realist interpretations on the debate and argue that they also imply eternalism. I then conclude that all the philosophical interpretations of the non-fundamentality of time entail a form of eternalism.

\section{Presentism, No-Futurism, Eternalism}

According to presentists, present entities possess the ontological privilege of existence: past and future entities do not exist \textit{simpliciter} (see e.g. \citealt{bigelow1996presentism}, \citealt{markosian2004defense} and \citealt{bourne2006future}). For eternalists (see e.g. \citealt{Smart1963}, \citealt{mellor1998} and \citealt{sider2001}), past, present and future entities exist \textit{simpliciter}. No-futurism---also called past-presentism or the growing block view when regarded in conjunction with realism about the flow of time, tensed facts or another category of entities conveying the idea of a primitive becoming and of a change in the material content of reality depending on the time of reference---states that past and present, but not future, entities exist \textit{simpliciter} (see e.g. \citealt{broad1923scientific}, \citealt{tooley2000} and \citealt{button2006there, button2007every}).\footnote{I am aware that the way I introduce the distinction between these views will sound problematic for many. On the one hand, philosophers of physics may complain that it does not take into account what we learn from General Relativity, in particular that we should avoid any reference to past, present or future entities, using instead the notions of anteriority and posteriority in a reference frame to define eternalism. On the other hand, metaphysicians might argue that we may approach the problem differently, for instance by avoiding reference to existence \textit{simpliciter}. See e.g. \cite{Crisp2004-CRIOPA} and \cite{deasy2017presentism} for discussion. However, the old-fashioned classification I use is helpful to introduce, as a first go, how we think intuitively of those views.} The term \textit{simpliciter} is here to show that (most) proponents who accept that the debate is genuine appeal to the notion of existence \textit{simpliciter}, distinct from the notion of existence at a time. For instance, Platonism about mathematical entities, namely the view that mathematical entities do exist, is usually regarded as the view that mathematical entities exist \textit{simpliciter}---and not in time at a particular temporal location or everywhen. The debate about existence in time, likewise, begins with the assumption that it is meaningful to ask whether past and future entities exist \textit{simpliciter} or not. Importantly, all the players in the game agree that it is true that past entities do not exist anymore, and that future entities do not exist yet. What they disagree about is whether from the claim that past entities do not exist anymore we should conclude that past entities do not exist \textit{simpliciter}, or whether we should conclude that past entities do exist \textit{simpliciter}, although at another temporal location (likewise for future entities).\footnote{A few philosophers of physics such as \citet{dorato2006irrelevance} have argued or expressed the view that the whole debate lacks substance. Following \citet{wuthrich2012demarcating}, I will assume within the scope of this paper that temporal existence is an important and meaningful matter worth investigating.}


Eternalism and presentism enjoy symmetric advantages and drawbacks. Eternalists may easily account for the truthmakers of past statements. Some parts of spacetime, including their material content, act as truthmakers for these statements. For example, if I say ``Yesterday, I was in Chicago'', this statement is true because there is a spatio-temporal part of reality---the volume of spacetime corresponding to the day of yesterday in Chicago--—that makes it true. But it seems difficult for the eternalist to account for the intuition that future contingent facts are unsettled. Indeed, by holding that future entities exist, we have to face a threat of \emph{existential determinism}: if the material content of spacetime at the date of tomorrow (i.e., in a relativistic setting, a few hours from here in my future light cone) is already there, meaning that it exists over in the future as from here, the contingency of these events seem to be lost. For instance, if I say ``tomorrow, I will be in Oxford'', this statement is already true now, because the fact making it true exists as a part of space–time.\footnote{But see \cite{BLBphd, BLBV} for a defense of the claim that eternalism is consistent with an open future.} Presentists have to deal with symmetric drawbacks and advantages: they may explain why contingent future facts are unsettled---it is just because the future does not exist. But they have a hard time finding truthmakers for past statements since the past does not exist \textit{simpliciter}. On the positive side of things, the presentist can easily accounts for the fact that ``tomorrow I will be in Oxford'' does not have a truth value yet: it is just that it does not have a truthmaker. It thereby accounts for our ordinary intuition that the future is not (at least completely) fixed. On the negative side of things, by refusing the existence of the world at the date of yesterday, statements such as ``yesterday I was in Chicago'' are threatened to lose their truth, and the past is at risk of loosing its fixity.

When it comes to relativity, it is important to notice that several attempts have been made to reconcile presentism with special and/or general relativity, such as the neo-Lorentzian reinterpretation of special relativity \citep{Craig2001-CRATAT-4}, Fine's fragmentalist approach to special relativity \citep{Fine2005-FINTAR-2} and approaches looking either at cosmology or quantum gravity to find an objective foliation of spacetime \citep{monton2006presentism}. However they all fail for different reasons that I cannot review here. For a general overview and why ``whatever presentism remains compatible with empirical facts and our best physics is metaphysically repugnant'', see \cite{wuthrich2013fate}. For an explanation of why the parameter called cosmological time to be found in the standard model of big bang cosmology, based on general relativity, does not help recovering a global foliation of spacetime that the presentist may rely on, see \citet[section 5]{wuthrich2013fate} and \cite{smeenk2013time}.\footnote{Another radical option is to replace GR by another theory empirically equivalent to its relevant sector---the set of solutions that could describe the actual world---and that would rely on a different ontology. This is the program behind \emph{shape dynamics} (see e.g. \citealt{gomes}).}

No-futurism may sound appealing as a third way. By positing that past and present---but no future---entities exist, no-futurism could, \textit{prima facie}, inherit of the advantages of both eternalism and presentism---without their drawbacks. It is true that yesterday I was in Chicago because the part of spacetime at the date of yesterday exists \textit{simpliciter} and includes myself (or a temporal counterpart of myself, or a temporal proper part of myself). But the fact that I will be in Oxford tomorrow is unsettled because there is no such thing as a volume of spacetime corresponding to the date of tomorrow and including the relevant content. However, no-futurism conflicts with general relativity, in the same way presentism does. The no-futurist will need an objective slice of spacetime corresponding to the border of being; but according to general relativity, there is no objective foliation of spacetime in spatial layers.  

In the next section, I examine the potential disappearance of time in the two leading approaches to quantum gravity: \textit{string theory} and \textit{loop quantum gravity}.


\section{Time in String Theory}

The status of time in the most popular approach to quantum gravity, namely \emph{string theory} (ST hereafter), remains unclear. Indeed, there does exist an influential ontological interpretation of ST rejecting the existence of time---and as such, although ST is usually set up with a ``full-blooded time'', as Callender puts it, time may be regarded as a merely useful theoretical ladder that can be kicked away after it has been climbed. According to this \textit{two-dimensional field ontology}, endorsed by one of the leading proponents of ST, Edward Witten, spacetime---including both spatial and temporal aspects---does not exist (see \citealt{witten1996reflections}, reprinted in \citealt{callender2001physics} and \citealt*[section 3]{huggett2013time}, \citealt*{huggett2015deriving} and \citealt[section 4.2]{baker2016} for philosophical discussion). As Witten puts it:

\begin{quote}
...one does not really need spacetime any more; one just needs a two-dimensional field theory describing the propagation of strings. And perhaps more fatefully still, one does not have spacetime any more, except to the extent that one can extract it from a two-dimensional field theory. \citep[28]{witten1996reflections}
\end{quote}

So what are those two-dimensional fields exactly? To understand this, let us sketch out the ontological picture behind ST.\footnote{For the technically-minded philosopher: I begin by sketching perturbative superstring theory---I will then comment on what might be expected, ontologically speaking, from a non-perturbative M-theory.} Strings are one-dimensional spatial entities extended through time, and one may regard any perduring string as a \emph{two-dimensional worldsheet}.\footnote{The ontology of ST sometimes also includes branes which are \textit{n}-dimensional entities extended through time, giving rise to \textit{n}+1-dimensional geometries. In this two-dimensional field approach, branes are not among the mereological building blocks of the world.} One might then define an \emph{internal spacetime}---corresponding to the two-dimensional worldsheet swept out by the strings---living in a ten-dimensional background spacetime. Depending on how we think about the measurement problem and the notion of quantum superposition at the worldsheet level, we may end up with an ontology with superposed internal spacetimes (relying on a many-worlds approach to this superposition) requiring a many-to-one connection between the fundamental worldsheet\textit{s} and one and only one emerging relativistic spacetime, or a many-to-many connection between a quantum superposition of worldsheets and a quantum superposition of relativistic spacetimes. This depends on whether emergence should play a role or not in the quantum-classical transition and how we should analyze superposition in the first place.\footnote{Also, one may ask whether the picture actually makes sense in case strings are not interacting, leading seemingly to various disconnected internal spacetimes. This has been pointed out by Keizo Matsubara according to \citet[10]{baker2016}. This is an interesting issue for the two-dimensional approach but not necessarily an unpalatable one. Indeed, it may well be that the actual world is made of causally disconnected islands (see e.g. \citealt{bricker2001}).} Each string has a few properties. Some of these properties, namely vibrations, correspond to physical particles. Note at this stage that we do have \emph{two structures} that could be called space or spacetime within the ontology of ST: the \emph{internal two-dimensional spacetime} associated with the collection of strings fusing and splitting and the \emph{ten-dimensional background spacetime} in which they live.

Now, we may ask to which, if any, of those two structures corresponds \emph{GR spacetime} and, by extension, \emph{space and time as we observe them around us}---insofar as space and time as we ordinarily perceive them may be regarded as local abstractions from GR spacetime. In order to ease the discussion, let us call, following \citet{HuggettDualities} and a particular use in the philosophy of quantum gravity community, the conjunction of GR spacetime and ordinary space and time ``phenomenal'' spacetime---which should not be equated with \text{phenomenological} space and time as usually understood in the broader philosophy community, namely to space and time as we ordinarily experience them. The question of importance for the metaphysician of time is then: \emph{Should we identify phenomenal spacetime with internal spacetime, background spacetime or neither of them?} In other words, what is the relevant structure in the ontology of ST for addressing metaphysical questions about the nature of time?

Let us consider the perhaps most obvious answer to this question: \emph{phenomenal spacetime is the background spacetime} in such a way that string theory simply asserts that physical spacetime is made of ten, and not four, dimensions, one dimension being time. If this were to be the case, then ST would not be saying anything novel with respect to GR about time. However, there exists a deep symmetry in descriptions of the background spaces, the so-called ``target spaces''--namely, T-duality---which renders background space with different compactification radii \emph{physically equivalent}. A natural \emph{common core interpretation}\footnote{There do exist other ontological interpretations of duality (see e.g. \citealt{BLBJR}). But what matters for our purpose is that, as \citet{HuggettDualities} argues, all interpretations start with the claim that we have no reason to identify GR spacetime with \textit{one} dual background structure alone, barring the road to identifying GR spacetime with \textit{any} background structure.} of dual descriptions then leads to the claim that this structure is not identical to phenomenal spacetime, which does have a determinate compactification radius. This \emph{T-duality seem to imply that background space is not real and makes it a poor candidate for identification with phenomenal space}. However, this duality afflicts background space only and not background time---so in this view time remains fundamental and the issue is only about the emergence of GR spacetime from the background space.

What about \emph{internal spacetime}? Could we identify time in this picture as the internal time within the worldsheet? This is the two-dimensional field proposal: the worldsheets, interacting and composing the global complex worldsheet, are regarded as the \emph{building blocks of the physical world}. Instead of thinking of the strings as entities evolving in the background spacetime, in this picture the so-called background structure is regarded not as spacetime, but as a \emph{ten-dimensional vector field} living on the two-dimensional worldsheet (for a more technical presentation, see \citealt*[251-252]{huggett2013time}). The world would literally be made of two dimensions at the fundamental level, in such a way that most of the complexity we observe at our scale would emerge from the very complex fluctuations of fields located on the worldsheet. So what is the nature of time if internal spacetime is the closest structure to phenomenal spacetime? As \citet*[section 3]{huggett2013time} explain, this internal time is subject to \emph{conformal symmetry} and thereby to \emph{scale invariance}: attributions of durations between points within the worldsheet is a quantity that can be chosen arbitrarily, without making any physical difference in the worldsheet.\footnote{More technically: ``the action is invariant under arbitrary local rescaling of the string metric'' \citep*[254]{huggett2013time}.} What it means is that although \emph{spatio-temporal ordering} is still implemented in this picture, there is no matter of fact about definite \emph{temporal distances} (i.e. durations) obtaining between couples of events (i.e worldsheet points).\footnote{It could be objected that a non-perturbative string theory might change the fact that the two-dimensional field approach is a natural ontology for ST. It might indeed. However, first we do not have such a theory at the moment, and second M-theory might still be consistent with the two-dimensional ontology, which, after all, is supposed to approximate M-theory. For those two reasons it is worth investigating the ontological consequences of this view.}






\section{Time in Loop Quantum Gravity}

Let us now look at \emph{loop quantum gravity} (LQG here after), one of the most promising of these research programs denying the fundamental existence of time. According to LQG, what there is instead of spacetime are entities (or a structure) \emph{described by} ``spin networks''---namely, collections of nodes and relations (the loops) between these nodes (see \citealt{rovelli2004quantum} and \citealt{RovelliVidotto}; for a summary aimed at philosophers cf. \citealt[279-280]{HuggettWuthrich2013}). The 3D spin networks representations can be replaced by 4D representations called ``spin foams'', when we apply a dynamic to the spin networks. Spin foams are conjectured as sufficiently similar to GR spacetime to explain the empirical success of GR. Spin foams describe discretely valued volumes and areas prompting a question about the relation between these discrete structures and the continuous GR structure. But this is not the most important point. Importantly, this description of the structure in terms of discrete geometrical entities should not be conflated with the \emph{fundamental structure of the world}; it should rather be regarded as a useful way to represent it and handle its quantum states (cf. \citealt{RovelliVidotto} and discussion by \citealt{wuthrich2017raiders} and \citeauthor{healey} forthcoming, section 3). The underlying ontology is regarded as not including spacetime (discrete or not).\footnote{See \citet{norton2015loop}.}

At the moment, I believe it is fair to say as \citeauthor{healey} (forthcoming) does, that there is no consensus among physicists about the ontological nature of the fundamental LQG structure lurking behind the mathematical spinfoam structure. I propose to distinguish between two approaches in order to make sense of this fundamental structure. Either this fundamental structure has enough spatio-temporal structure to be described as a \emph{quasi-spacetime} or it does not. According to the first approach, the fundamental structure is a ``quasi-spacetime'' which is very different from the ordinary spacetime---namely, spacetime as we theorize it with GR. But this structure also looks similar, to some degree, to the ordinary, phenomenal, spacetime by sharing properties with GR spacetime (for instance by being made of concrete physical relations organized in a network). According to this approach, this structure shares the metaphysical profile of GR spacetime by belonging to a \emph{general category of ``physical spacetimes''}. According to the second approach, the LQG structure should rather be regarded as \emph{a structure that is not spatio-temporal at all}. The LQG structure is a sort of physical structure hard to categorize ontologically that can only be described in purely algebraic terms. This approach entails that it is impossible to make sense of the LQG structure in terms of the metaphysical categories used in general to conceive of the spatio-temporal natural world (relations, properties, events, etc.) and that the content of the black box should not be analyzed at all in terms of those metaphysical categories.

Importantly for our purpose, and independently of the way we think about the status of the fundamental structure, the organization of the LQG structure may not correspond systematically to the spatio-temporal ordering of events, as described by GR: indeed, in some models of LQG, some \emph{relations of adjacency} in the LQG structure correspond to \emph{relations of large spatio-temporal distances} in the GR structure. See e.g. \citet{markopoulou2007disordered} and \citet[279-280]{HuggettWuthrich2013} for a philosophical discussion.\footnote{Disordered locality remains speculative for two reasons. First, disordered locality has been shown to apply to space, in some models. One should expect disordered locality to apply also to timelike separated events, but this is a conjecture. Second, since we do not have any fully worked out theory of LQG, we cannot exclude that disordered locality in general might be linked to a mathematical artifact with no physical content.} It entails in particular that it happens that what we regard as large temporal distances in GR correspond to relations of adjacency in LQG. And here it is important to see that this point holds independently of the attitude we take regarding the nature of the fundamental structure. What matters here is not whether the black box is ``metaphysically empty'', being only described by algebraic mathematics, or whether we can attribute an internal metaphysical structure to this physical structure in terms of, say, relations and nodes. What matters is that the way we must ascribe quantum states to the black box---independently of what there is in the black box---deviates from the system of locality described in GR. Therefore, what comes under attack with the non-existence of fundamental time in LQG is not only the shift from a discrete structure (in LQG) to a continuous structure (in GR), or from the quantum to a classical world. The very existence of the \emph{partial ordering of events} as we observe them in daily life, and conceptualize them through GR, might only be an approximation at our scale of what is really the case.




\section{Consequences for Existence in Time}

As a result, if further work shows that the right ontological interpretation of ST does not include time, then we will end up with a situation in which the two most popular approaches to quantum gravity assert that time is absent from our most fundamental description of the natural world. This would give us strong evidence in favor the non-fundamentality of time along the ambitious epistemological path that I mentioned in the introduction. I will not take a stance on this though; within the scope of this paper I merely point out, along the modest epistemological path, that \emph{some} approaches to quantum gravity entail the non-fundamentality of time.

This denial of the \emph{partial ordering of events} in LQG, and of \emph{duration} in ST, is more radical than what we usually find in the recent philosophy literature. Indeed, in the metaphysical and phenomenological literature, it is quite common to argue that time does not flow (the quite standard B-theory), and that the notion of flow corresponds to a perceptual or linguistic artifact (see for instance \citealt{paul2010temporal}, \citealt{frischhut2015experience} and \citealt{benovsky2015experience}). Also, it is sometimes argued that time does not own an intrinsic direction (the C-theory according to which the temporal dimension is only made of non-orientated relations). In a relativistic setting, these claims amount respectively to the claims that spacetime is not flowing along a temporal direction, or that spacetime does not own an intrinsic direction.  However, it is almost universally agreed that although spacetime is less temporal than what is commonly thought, there does exist a \emph{local split} between space and time in the actual world (see e. g. \citealt[chap. 6]{callender2017makes}).\footnote{For any four-dimensional volume in spacetime, the temporal dimension differs from the three spatial dimensions with respect to several features. To give a simple example, GR is locally special relativistic in the sense that one can always pick a local coordinate system such that at its origin: (1) the metric takes the form of the Minkowskian metric, and (2) the dynamical matter equations remain invariant under Lorentz transformations. This in turns also means that the SR-distinction between space and time is inherited at least locally. But note that our actual world distinguishes space and time in various, and seemingly logically independent, ways: Callender names this feature the ``fragmentation of time'' \cite[chap. 6]{callender2017makes}.} 



How does the non-fundamentality of time, in the LQG and ST contexts, compare with these metaphysical claims made against the existence of some features usually ascribed to time? Those approaches do not focus on dispensable properties belonging to the rich conception of time such as flow and intrinsic directionality but with \emph{bulk properties} of time: \emph{ordering} and \emph{temporal distances}. However, these attacks on bulk properties of time are not as radical as they could be because these approaches preserve a distinction between quasi-spatial and quasi-temporal structures, namely a local split between quasi-space and quasi-time, like in GR (see \citealt[section 3.2]{BLBNL}). This situation leads to a subtle verdict: the two most promising approaches to quantum gravity entail the non-fundamentality of time if duration and/or GR partial ordering are regarded as essential properties of time or, in other words, as necessary conditions for time to obtain. However, if the only necessary condition for time to obtain is the possibility to distinguish locally in the structure a set of directions, then LQG and the two-dimensional field approach to ST do not entail the non-fundamentality of time. Following \cite{BLB2015unrealities}, I will assume that this question is a merely linguistic matter, and will focus on a more substantial question: in the two cases at hand, do we have \emph{enough of time} in order to express the classical views of existence in time?



In the context of LQG, disordered locality conflicts with presentism, no-futurism and, we shall see, implies a particular understanding of eternalism. Presentism and no-futurism rely on the \emph{conceivability of a global separation between three zones of reality}: the past, the present and the future. In GR, this global separation needs to be forced on the model, which does not offer any reason to believe that there is such a thing. But let us admit that this is so for the sake of argument. Say that [\textit{a},\textit{b},\textit{c},\textit{d},\textit{e}] is an ordered collection of timelike separated events, causally connected on a particular timelike worldline in the GR description and that LQG describes those entities as: \textit{a'},\textit{b'},\textit{c'},\textit{d'},\textit{e'}. Now let us suppose that we are localized at \textit{c} and that there is an objective foliation individuating a special slice of spacetime including \textit{c}. This \emph{imposed foliation from the macroscopic level will appear in the fundamental LQG description to be in conflict with disordered locality}. For instance, although event \textit{b} is causally before \textit{c}, and event \textit{d} is causally after \textit{c} in the GR description, disordered locality entails that it sometimes happens in the LQG description that \textit{c'} is \emph{not between} \textit{b'} and \textit{d'} (because of the existence of a different ordering, and excluding the case of spacelike separated events). Therefore, the macroscopic foliation used to distinguish between objective past, present and future zones of the macroscopic spacetime can only be a statistical approximation, obtained by neglecting disordered locality. But if the macroscopic foliation is a statistical approximation then it does not seem right to take it to carve nature at its joints. This is an issue for presentism and no-futurism since a macroscopic foliation carving nature at its joints is a preliminary condition required in order to define those views. What about eternalism? As eternalism does not need relying on a genuine metaphysical distinction between the past, the present and the future (namely, on an A-theory), it may avoid this issue. However, even the B-theoretic eternalist view---namely, the view that all natural things exist independently of their location in the network of spatio-temporal or causal relations---is threatened to some degree by disordered locality: if the causal ordering of events, which is part of the ontology according to the B-theory, is only a statistical approximation then the very core causal structure of the four-dimensional world is emergent and may not be used to define eternalism.



Therefore, disordered locality requires adopting another view about existence \textit{simpliciter} in the natural world. I call this view ``atemporal eternalism'': all proper parts of the natural world exist \textit{simpliciter} and the material content of the natural world does not depend on any particular location in it. This timeless view resulting from disordered locality may sound like standard eternalism when it comes to existence, and indeed it is very similar to it. In fact, I take atemporal eternalism to clarify the fact that eternalism is only superficially tied to the existence of time. What matters with eternalism is not that all things equally exist in time, but rather than all things that are parts of the natural worlds equally exist. This core eternalism can be then complemented with the view that the location system that allows us to target different parts of the natural world should be formulated in terms of time or spacetime locations in a spatio-temporalist framework---or, on the contrary, in a non-spatio-temporalist ontology with a non-spatio-temporal system of location. Indeed, whether or not ``atemporal'' eternalism is just standard eternalism well-conceived is a terminological dispute that may be given various reasonable answers. What matters here is that the new eternalist account---independently of whether one wants to classify it as a form of standard eternalism, or as a new kind of eternalism---is more radical by preserving the core idea behind eternalism, namely that existence is not restricted to our immediate spatio-temporal surroundings\footnote{One might argue that this expression of ``surroundings'' does not make sense in this context. But if so, then it strengthens the fact that we are not dealing with standard eternalism here. Furthermore, whether this expression does or does not make sense here depends on whether the fundamental LQG structure should be identified to a \textit{quasi-space} or not. If it is a quasi-space, then it should in principle be possible for this structure to have enough geometrical features to use meaningful quasi-spatial predicates modeled on spatial predicates.}, without committing further to the claim that there exists some way, either in A-theoretic terms or in B-theoretic terms, to slice in a non-approximative way the four-dimensional cosmos into 3D hypersurfaces. In other words, according to atemporal eternalism, \emph{any physical entity, independently of its location in the fundamental structure, exists} simpliciter \emph{and does not depend for its existence on any vantage point in the structure}. There is no change in the material content of reality because there is no objective time, and reality---taken as a whole---does not change. Compare this with standard eternalism: relatively to any time \textit{t}, all other entities located in time at other instants do exist \textit{simpliciter}---since existence \textit{simpliciter} does not depend on the temporal location within the temporal structure. Or, in a relativistic framework, relatively to any part of spacetime, all other parts of spacetime, including their material content, exist \textit{simpliciter}. In standard eternalism, we find a reference to time---if only to claim that this reference does not constrain what exists---when in atemporal eternalism such a reference is missing. But, importantly, the two views agree that the range of existence \textit{simpliciter} does not depend on any vantage point in the (temporal or atemporal) structure: existence is not a \textit{local} matter. In other words, independently of the location in the non-spatio-temporal structure, the scope of what there is will range over the totality of physical entities existing in, or constituting, the physical structure. As such, atemporal eternalism is faithful to the spirit of standard eternalism but goes one step further in the denial of the classical terms of the debate, by rejecting the possibility to distinguish locally (in A-theoretic or B-theoretic terms) between zones of time.

From the claim that the two most promising approaches to quantum gravity entail the non-fundamentality of time and thereby, as we just saw, (standard or atemporal) eternalism, we may conclude along the \emph{moderate epistemological path} that we have substantial, but no absolute, evidence in favor of the truth of some sort of eternalism.\footnote{The perhaps most famous approach to quantum gravity that, it has been claimed, is consistent with---or even suggests---a growing block interpretation is the \textit{causal set theory} (see e.g. \citealt{rideout1999classical}, \citealt{dowker2006causal}, \citealt{dowker2014birth} and \citealt{earman2008reassessing}, \citealt{HuggettSkeptic}, \citealt{Wuthrich}, \citealt{wuthrich2016becomes},  \citealt[section 5]{callender2017makes} for a philosophical discussion). Although, I cannot go fully into detail here, I simply want to point out that in this approach the growth is only \emph{local}. As \citet[100]{callender2017makes} puts it: ``[the] `birthing' process is said to unfold in a `generally covariant' manner''. There is no unique way to group together different causal sets that are not causally connected and to take a stance on which of them do exist as of the perspective of a particular causal set. It simply is impossible to quantify over what exists in other causal sequences as from the perspective of a particular causal set in this picture. Therefore, this growing view is not the classical growing \emph{block} package including the no-futurist component. This view could more adequately be described as a ``\emph{growing octopus view}'' since one might think (metaphorically) of all the various causally disconnected growths as octopus' tentacles extending independently. (Note that the metaphor is not completely accurate since the tentacles should split again and again into novel tentacles.)} 



\section{Eliminativism, Reductionism and the Derivative View}


Let us now turn to the question of what it means that time (meaning, depending on the context, duration or temporal ordering) does not \emph{fundamentally} exist. Indeed, one may argue that although presentism, no-futurism and standard eternalism may not apply at a fundamental level of description, under certain hypotheses it may nonetheless be the case that these views may correctly describe other scales of descriptions. In order to examine this possibility, I will distinguish between \emph{three conceptions of the non-fundamentality of time}. The three interpretations follow closely a classification made by \cite{LeBihan}, \cite{le2018space}, \cite{BLBNL} and \citeauthor{BLBsynthese} (forthcoming) about the non-fundamentality of \textit{spacetime}. As we shall see, the classification works as well in the case of the non-fundamentality of \textit{time}. 

In the philosophy of mind, we may distinguish between different issues: \emph{easy problems} and the \emph{hard problem of consciousness}. The easy problems consist in finding the necessary and sufficient conditions for the existence of mental states, couched in scientific terms, plus explaining away intentionality. In contrast, the hard problem of consciousness asks about the correct ontological interpretation of the relation of consciousness with the physical world. What is the source of the correlations between mental states and physical states? Do we observe these correlations because the mental is identical with physical phenomena? Or do we have here distinct kinds of substances or properties related by a relation of some kind? Or does it mean that the mental is a pure illusion and does not exist? With consciousness, because of the qualitative aspect of our experience, the ``what it is like to experience'', and its subjective aspect, the fact that these experiences seem to point in the direction of a subject having these experiences, we develop a rich terminology quite hard to translate in the vocabulary of physics. It seems that we have a mystery to deal with, namely the presence of an apparent explanatory gap between two terminological worlds that seem to be very different: the physical realm and the mental realm. This problem is said to be hard because it does not seem that science could help us to solve it. It appears to be quite abstract, perhaps rooted in the concepts we use to describe the world.

A similar \textit{hard problem of spacetime} appears to some extent in the context of spacetime emergence: it seems that offering a formal derivation of a spatio-temporal theory from a non-spatio-temporal theory is not enough to make sense, ontologically speaking, of the apparent conceptual discrepancy between the primitive notions of the spatio-temporal theory, and the primitive notions of the non-spatio-temporal theory (see \citealt{maudlin2007completeness} for a discussion of the issue in quantum mechanics, \citealt{HuggettWuthrich2013} for a discussion of the issue in quantum gravity with an epistemological focus and \citeauthor{BLBsynthese} forthcoming for a clarification of the ontological issue). 

Likewise, the ``hard problem of \emph{time}''---in analogy with the hard problems of \emph{consciousness} and the hard problem of \emph{spacetime}---addresses the explanatory gap between a temporal theory and a timeless theory. The temporal vocabulary is deeply rooted in our conceptual apparatus, and it is far from clear how we should regard the reference of this vocabulary, metaphysically speaking, if time is not fundamentally real. What matters here is that, even if we manage to solve the ``easy'' problem in offering a derivation of general relativity and quantum field theory from the novel theory, a gap will still have to be filled: why is it the case that a temporal theory works as an approximation of a timeless theory? As with the hard problems of \emph{consciousness} and \emph{spacetime}, I suggest to divide the possible answers to the hard problem of time in three categories: \textit{dualism}---that for a reason that will become clear, I will call the \textit{derivative time view}---\textit{reductionism} and \textit{eliminativism}.
	
Dualism is the view that both the \emph{fundamental timeless structure} and the \emph{derivative temporal structure} are real. The term ``dualism'' entails the existence of two categories of entities, but it does not say much about the relationship obtaining between the two categories. In particular, it remains silent on whether one collection of entities is \emph{more fundamental} than the other. The idea, here, is that the structure described by a forthcoming theory of quantum gravity, and the entities this structure is made of, are more fundamental than the structures and entities described by both GR and quantum field theory. Therefore, I will choose to refer to this view under the name of ``derivative time view''. According to the \emph{derivative time view} \citep{Wuthrich} time exists derivatively.\footnote{Wüthrich goes as far as defending that closed timelike curves might be excluded at the fundamental level while nonetheless obtaining at the derivative level (\citeauthor{timetravel} forthcoming).} The derivative time view requires to posit three things: first, layers of reality (i.e. ontological levels in contrast to merely descriptive, epistemological levels), second, a mind-independent ontological connection between the entities constituting these levels of reality, third, a more-fundamental-than relation obtaining between the entities making up these levels. This relation of fundamentality might be primitive or defined with respect to another notion, say a \emph{grounding relation}, or a \emph{building relation} (see \citealt{bennett2017making}).
	
According to the \emph{reductionist view} \citep{le2018space}, both temporal and timeless entities are real; however, it does not entail---\textit{contra} the derivative view---that reality is stratified in levels with the temporal level being less fundamental than the timeless one. Temporal, or more precisely for this matter, spatio-temporal entities can be regarded as being \textit{literally composed of non-spatio-temporal entities} by using the notion of \textit{logical composition} as introduced by \cite{paul2002logical, paul2012building}. The notion of logical composition was introduced in order to formulate the view that material objects (like chairs) are genuinely composed of their properties (say, their colors, shapes, and primitive locations). This shows that it is conceivable to regard the natural world as being structured of various metaphysical categories, and that composition need not relate entities that belong to the same metaphysical category. The reductionist view does not accept this mereological bundle theory (the view is at odds with the claim that spatial or spatio-temporal locations are primitive properties), but it posits that wholes and their parts need not belong to the same metaphysical categories. It is not clear how we should analyze the temporal realm and the non-temporal realm in terms of metaphysical categories, but the point is that, whatever the relevant analysis turns to be, we may regard the connecting device between the two realms as being some sort of composition. Like the derivative spacetime view, the reductionist view acknowledges the existence of both temporal entities and timeless entities. And it posits a mind-independent relation obtaining between the two kinds of entities.


Therefore, the difference between the derivative and the reductionist views is that the first states that the notion of composition cannot account for spacetime emergence, when the second identifies spacetime emergence with a particular case of composition, stressing that composition need not always be spatio-temporal. The reductionist view, by selecting logical composition as a primitive notion to do the trick of relating the spatio-temporal to the non-spatio-temporal avoids ontological anti-reductionism, namely stratifying reality into ontological levels.



What about \emph{time eliminativism}? According to eliminativism, only the structure described by the fundamental theory exists. Spacetime is not fundamentally real because it does not exist \textit{simpliciter}. In this ontology, there is no need for levels of reality, trans-level connecting devices and more-fundamental than-relations. Although this approach might seem metaphysically appealing, it makes the problem of empirical coherence \citep{HuggettWuthrich2013} more salient by denying that evidence justifying our theories are localized in some non-fundamental spacetime. Wüthrich brings our attention to this point when he writes:

\begin{quote}
[It] is a necessary condition for an empirical science that we can at least in principle measure or observe something \textit{at some location at some time}. The italicized locution, in turn, seems to presuppose the existence of space and time. If that existence is now denied in quantum theories of gravity, one might then fear that these theories bid adieu to empirical science altogether. It thus becomes paramount for advocates of these theories to show that the latter only threaten the fundamentality, but not the existence of space and time. \citep[298]{wuthrich2017raiders}.  
\end{quote}

I take these three options to be exhaustive: the emergence of time means that time derivatively exists, reductively exists, or does not exist.\footnote{Explaining emergence in terms of reduction might come as a surprise for the reader mainly trained in philosophy of mind or philosophy of emergence since emergence is often defined negatively as an impossibility to reduce a terminological framework to another one. But note that in physics and philosophy of physics, it is common to regard emergence and reduction as consistent tenets since emergence is a broader term (see e.g. \citealt{butterfield2011less}).} I will not take position on these possible interpretations here and will rather examine their consequences for existence in time.\footnote{\citet{Braddon-MitchellMiller} recently claimed that only the eliminativist interpretation of time emergence is worthy of philosophical discussion. They write:

\begin{quote}
Let us distinguish between what we call weakly timeless theories and strongly timeless theories. Weakly timeless theories are theories in which although there is no time at a fundamental level, time, or something very time-like, emerges at a macro level. [...] In what follows we shall not be interested in weakly timeless theories, since although it is an interesting discovery that time is emergent, rather than fundamental, no particular philosophical problems arise from such theories, and it is not really clear that calling these theories timeless is appropriate. \citep[1808]{Braddon-MitchellMiller}
\end{quote}

Their claim is puzzling. They merely consider the possible emergence of time within a particular physical theory and do not consider the emergence of a temporal physical theory from a distinct physical timeless theory. Explaining how such an emergence is possible constitutes a highly interesting philosophical issue.}

The consequences of eliminativism for existence in time are self-explanatory: the view entails standard eternalism in the case of the two-dimensional field approach to ST and atemporal eternalism in the case of LQG. But the possibility that time might emerge in the sense of being derivatively or reductively real leads us to a new question: could it be that one of the classical views about existence in time is correct about the macroscopic level but not about the fundamental level?

\section{From Realist Views to Eternalism}

In this section, I argue that if the derivative time view, or time reductionism, is the correct interpretation of time emergence, then (standard or atemporal) eternalism is true. When we accept that time emergence should be interpreted along a stratified picture, time being derivatively or reductively real, how should we understand the debate over existence in time? To begin with, according to time realism (namely, the disjunction of the derivative view and the reductionist view), \emph{both the fundamental\footnote{For the ease of presentation, in what follows, I will use the expression ``fundamental ontology'' and refer to entities being more fundamental than other entities, assuming that the word ``fundamental'' should be interpreted differently depending on whether we are operating under the assumption of the derivative view (then we have a notion of fundamentality which is not identical with mereological composition) or time reductionism (then the more-fundamental-than relation is just logical parthood).} non-spatio-temporal and the spatio-temporal structures are real}. This approach may prompt two questions. 

First, is the \emph{fundamental ontology} presentist, eternalist or no-futurist? As I have argued in the previous section, according to LQG and ST, the fundamental structure should be interpreted as being eternalist. Second, could the (derivatively or reductively) real temporal structure be regarded as described by presentism, no-futurism or eternalism? In other words, could it be that, although the world is, say, atemporally eternalist at the fundamental scale of description, the world might display at other scales---and in particular our macroscopic scale---a different temporal ontology such as presentism, no-futurism or standard eternalism? In response to this question, it could be argued that it would be somewhat strange to claim that \emph{different temporal ontologies} might obtain at \emph{distinct levels of reality}. If standard or atemporal eternalism holds true at the basal level of description, i.e. the one accurately described by a quantum theory of gravity, perhaps the same should be the case at higher levels of description. In other words, the natural world should satisfy, one might perhaps claim, a \emph{principle of trans-level metaphysical homogeneity}. According to this principle, ontology must be the same at all levels of description: entities must fall under the same primitive metaphysical categories and satisfy the same metaphysical principles. For instance, if properties are \emph{universals} then properties are universals at all ontological levels, at the fundamental and at any other level. Properties do not suddenly become, at a particular scale of description, \emph{tropes} or \emph{classes of objects}. Or, to take another example, if the fundamental ontology is one of \emph{objects}, then the derivative ontology cannot be an ontology of \emph{facts}. If trans-level metaphysical homogeneity is true in general, then it is true in particular for existence in time. Such a principle is \textit{prima facie} appealing by being grounded in considerations of \emph{ontological parsimony}: why multiply the ontologies within the actual world if it is possible to account for philosophical issues with the same ontology at all descriptive levels?

However, I believe this line of thought should be discontinued for at least three reasons. First, this principle already comes under attack with the derivative time reading: by definition, time (or, more precisely, spacetime) is real at the derivative level but not at the fundamental level. Therefore, operating under the assumption of the derivative time view is already a departure from the principle of trans-level metaphysical homogeneity. In other words, subscribing to the derivative view amounts to opening Pandora's box by rejecting the universality of metaphysical homogeneity---undermining the possibility to use it blindly as a systematic methodological guiding principle. 

Second, one might argue that one simply does not share this intuition and subscribe to \emph{ontological pluralism} according to which the ontology differs at various scales of descriptions. This move is quite common when it comes to defending a form of theoretical anti-reductionism for special sciences such as molecular biology or quantum chemistry grounded in some ontological anti-reductionism---namely, the conjunction of ontological pluralism and of an indexation of the various ontologies to scales of description. And if in this case, differences about what exists generate classes of contradictory facts about what exists (we will discuss this with more detail below), one should take note that it has been argued on several occasions that it might well be that the world is made of contradictory facts constituting \emph{temporal fragments} (cf. \citealt{Fine2005-FINTAR-2}) or \emph{dual fragments} (see \citealt[section 5]{BLBJR}).

Third, \emph{intuitions based on phenomenological experience alone cannot vindicate any philosophical claim about the ontology of scales which remain out of phenomenological reach}. Our phenomenal intuitions arose in a particular setting---namely, a particular level L*, and it would be premature to extend the ontology of L* to all Ls on this base alone. To put it metaphorically, we may well have evolved in a monochrome ontology which is not representative of the colorful richness of distinct ontological systems constituting reality---and obtaining at different ontological levels. 

For these three reasons, we need to carefully examine the possibility of \emph{scale fragmentalism}, namely that different ontologies might obtain at different scales of description, as a way to combine the non-existence of time at the fundamental level with one of the three temporal ontologies at human scale. 

I will now offer a specific argument against scale fragmentalism \emph{for the particular case of temporal ontology}. Let us assume that atemporal eternalism accurately describes the fundamental structure, but not the higher levels. The higher-level ontology is captured by one of the three other views: presentism, no-futurism or standard eternalism. The view entails an indexation of existence \textit{simpliciter} to particular levels. To put it differently, existence \textit{simpliciter} becomes relative to levels, and there is no existence \textit{simpliciter} \textit{simpliciter}. Relatively to a particular ontological level, the range of what exists \textit{simpliciter} differs (or can differ) from the range of what exists \textit{simpliciter} relatively to other levels. Therefore, it entails restricting the notion of existence \textit{simpliciter} to levels. Existence \textit{simpliciter} is existence \textit{simpliciter}-relatively-to-a-particular-level. We should thereby distinguish between two kinds of existence \textit{simpliciter}: relative existence \textit{simpliciter} (existence \textit{simpliciter} at a particular ontological level) and universal existence \textit{simpliciter} (trans-level existence \textit{simpliciter}). In the resulting view, relative existence \textit{simpliciter} (namely, temporal existence) may vary in scope from one level to another.



Let us take a venerable example: the event E of Caesar crossing the Rubicon. This event E is regarded as obtaining in the past as of our 21th century perspective. According to presentism, E does not exist \textit{simpliciter}. Now, according to the derivative time view, this may only be true at the derivative level and presentism should be reinterpreted as the view that \emph{E does not exist} simpliciter \emph{relatively to the higher level}. However, in the context of time emergence, E may be traced back to---i.e. associated with---a proper part of the fundamental non-spatio-temporal structure. Because eternalism adequately describes the fundamental structure, the grounding base of E in the fundamental structure exists \textit{simpliciter}, relatively to any perspective in the fundamental structure. Therefore, a proponent of derivative presentism would have to claim that it is both true that E \emph{does} and \emph{does not} exist \textit{simpliciter}. Although the grounding entity atemporally exists from the view point of the fundamental structure, the grounded entity does not exist \textit{simpliciter} as of the view point of the derivative structure. This relationalist line of thought (existence \textit{simpliciter} is indexed to levels) offers a way out for presentists and no-futurists.

However, it has the unpalatable aftermath of relativizing existence \textit{simpliciter} to levels, destroying the possibility to describe what there is in an \emph{absolute} way. Thus there is no definite answer to the question of whether Caesar exists \textit{simpliciter} since, in this picture, existential facts are to be relativized to scales of description. Furthermore, such a heavy theoretical framework (that requires subscribing to the derivative view against reductionism), by introducing so many ontological devices---ontological levels, distinct ontologies indexed to those levels and existential facts relativized to those levels---sounds rather undermotivated. Although scale fragmentalism allows, logically speaking, to defend derivative presentism or derivative no-futurism---some rather weak views since they concede that they do not apply at the fundamental level---this may only be done at the prohibitive cost of subscribing to radical ontological claims. So the only thing the proponents of presentism and no-futurism can do is to buy weak versions of their view at a tremendous cost. As a result, it is reasonable to assume that existence \textit{simpliciter} does not depend upon scales of description, with the implication that \emph{realist interpretations of time emergence (the derivative and the reductionist views) also entail standard or atemporal eternalism}.

\section*{Conclusion}
The non-fundamentality of time as we find it in loop quantum gravity and string theory, interpreted either in the eliminativist way or along some kind of realism, entails eternalism. Consequently, the dispute between presentism, no-futurism and eternalism may well be empirically settled in favor of standard eternalism, with the two-dimensional field approach to string theory, or atemporal eternalism, with loop quantum gravity.

\section*{Acknowledgments} For helpful comments on an earlier draft of this essay, I would like to thank Jiri Benovsky, Annabel Colas, Alberto Corti, Claudio Calosi, Fabrice Correia, Tiziano Ferrando, Vincent Grandjean, Rasmus Jaksland, Niels Linnemann, Cristian Mariani, Keizo Matsubara, Robert Michel and two anonymous reviewers. Special thanks to Nick Huggett and Christian Wüthrich for their invaluable feedback. This work was supported by the Swiss National Science Foundation.

\bibliographystyle{phil_review.bst}
\bibliography{references}
\end{document}